\documentclass{ws-ijmpa}

\begin{document}

\markboth{Z. Liu and Z. Chen}
{Quasilocal Non-Equilibrium Dynamics of $\phi$-Spinning Black Rings}

\title{QUASILOCAL NON-EQUILIBRIUM DYNAMICS OF\\ $\phi$-SPINNING BLACK RINGS}

\author{Zhenxing Liu}

\address{Wuhan Institute of Physics and Mathematics, Chinese Academy of Sciences,\\
Wuhan 430071, China\\
Graduate University of Chinese Academy of Sciences,\\
Beijing 100049, China\\
liunenu@tom.com}

\author{Zeqian Chen}

\address{Wuhan Institute of Physics and Mathematics, Chinese Academy of Sciences,\\
Wuhan 430071, China\\
zqchen@wipm.ac.cn}

\maketitle

\begin{abstract}
In this work, we study the non-equilibrium dynamics of $\phi$-spinning black rings\footnote{A
$\phi$-spinning black ring has rotation on the two-sphere $S^2$ rather than along the circle
$S^1$.} within the quasilocal formalism. We adopt the counterterm method and compute the renormalized
boundary stress-energy tensor. By considering the conservation of this tensor, the condition for
removing the conical singularity at spatial infinity is derived. It is subsequently shown that
a $\phi$-spinning black ring cannot be kept in a state of equilibrium, which is consistent with
the physical interpretation that the angular momentum is on the plane orthogonal to the ring
and there is no force to balance the tension and gravitational self-attraction. The results of
these computations lay a foundation for studying the thermodynamics of $\phi$-spinning rings
in detail. Finally, we charge up the rings in Einstein-Maxwell-dilaton system and suggest feasible
ways to make them balanced.

\keywords{black ring; conical singularity; quasilocal formalism.}
\end{abstract}

\section{Introduction}	

A remarkable progress in studying five-dimensional gravity is the discovery of a new object: the
black ring.\cite{1} Unlike the Myers-Perry black hole in five dimensions (see Ref. 2) which has
a spherical topology $S^3$, the black ring has an event horizon of topology $S^1\times S^2$.
During the past decade there has been growing interest in black ring physics motivated by looking
for novel properties not shared by four-dimensional black holes. For example, it was shown in
Ref. 1 that both the black hole and the black ring can carry the same conserved charges (specifically,
the mass and a single angular momentum), which implies the breakdown of conventional uniqueness
theorems in five dimensions.\footnote{However, there exist some five-dimensional analogs of the
uniqueness theorems, e.g., Refs. 3--5.} Many other interesting results can be found in Refs. 6 and 7.
By examining these results, we get a deep insight into the nature of gravity.

A considerable amount of effort has been devoted to the construction of new exact black ring
solutions in five dimensions. Supplemented with the counterterm, the quasilocal formalism of
Brown and York\cite{8} becomes a very powerful tool to investigate the properties of these
solutions. The conserved charges can be defined by using the divergence-free boundary stress-energy
tensor. There also exist all ingredients necessary to study in detail the thermodynamics of
various rings within this formalism.\cite{9}\cdash\cite{11} Furthermore, an important application
of the quasilocal formalism is to test the balance of a black ring. The balance condition results
from the conservation of the boundary stress-energy tensor.\cite{12}$^,$\! \cite{13}

The aim of this work is to carry out a study on the dynamics of $\phi$-spinning black rings
within the quasilocal formalism and to explore the mechanism for making these rings unbalanced.
Detailed computations and further discussions will be given in section 2 and section 3, respectively.

\section{Non-Equilibrium Dynamics}

We take the following metric\cite{14} as a starting point:
\begin{eqnarray}
ds^2=&-&\frac{H(y,x)}{H(x,y)}\bigg[dt-\frac{\lambda a y (1-x^2)}{H(y,x)}d\phi\bigg]^{\!2}\nonumber\\
&+&\frac{R^2}{(x-y)^2}H(x,y)\Bigg[-\frac{dy^2}{(1-y^2)F(y)}-\frac{(1-y^2)F(x)}{H(x,y)}d\psi^2\nonumber\\
&&\qquad\qquad\qquad\quad\ \ \,+\,\frac{dx^2}{(1-x^2)F(x)}+\frac{(1-x^2)F(y)}{H(y,x)}d\phi^2\Bigg],
\end{eqnarray}
where
\begin{equation}
F(\xi)=1+\lambda\xi+\left(\frac{a\xi}{R}\right)^{\!\!2}, \qquad H(\xi_1,\xi_2)=1+\lambda\xi_1+\bigg(\frac{a\xi_1\xi_2}{R}\bigg)^{\!\!2}.
\end{equation}
The parameters $R$, $\lambda$ and $a$ satisfy the inequality
\begin{equation}
\frac{2a}{R}<\lambda<1+\frac{a^2}{R^2},
\end{equation}
which eliminates closed timelike curves and guarantees the existence of event horizons. The coordinates
$x$ and $y$ have the ranges
\begin{equation}
-1\leqslant x\leqslant1, \qquad -\infty<y\leqslant-1.
\end{equation}
Asymptotic spatial infinity (we locate the boundary there) is reached as $x\rightarrow y\rightarrow-1$.

The conical singularity at $x=+1$ can be cured by identifying the angle $\phi$ with period
\begin{equation}
\Delta\phi=\frac{2\pi}{\sqrt{1+\lambda+a^2/R^2}}.
\end{equation}
However, another one still exists at $y=x=-1$. In order to remove it, we present a new method:
considering the conservation of the boundary stress-energy tensor\cite{9}
\begin{equation}
\tau_{ij}\equiv\frac{-2}{\sqrt{-h}}\frac{\delta I}{\delta h^{ij}}=\frac{1}{8\pi}\bigg(K_{ij}-
h_{ij}K-\Psi\Big(\mathcal{R}^{^{(4)}}_{ij}-\mathcal{R}^{^{(4)}}\!h_{ij}\Big)-h_{ij}\Box\Psi+\Psi_{;ij}\bigg),
\end{equation}
where $\mathcal{R}^{^{(4)}}_{ij}$ is the Ricci tensor of the induced boundary metric $h_{ij}$, $\Psi\!\equiv\!\sqrt{\!\frac{3}{\,\,2\mathcal{R}^{^{(4)}}}}$,
$K_{ij}$ is the extrinsic curvature of the boundary, and
\begin{equation}
I=I_{B}+I_{\partial B}=-\frac{1}{16\pi}\!\int_{B}\mathcal{R}^{^{(5)}}\!\!\!\sqrt{-g}\,d^5x-\frac{1}{8\pi}\!\int_{\partial B}\!\Bigg(K-\sqrt{\frac{3}{2}\mathcal{R}^{^{(4)}}\!}\Bigg)
\sqrt{-h}\,d^4x
\end{equation}
is the total action which has been renormalized by the counterterm\cite{15}
\begin{equation}
I_{ct}=\frac{1}{8\pi}\!\int_{\partial B}\!\sqrt{\frac{3}{2}\mathcal{R}^{^{(4)}}\!}\sqrt{-h}\,d^4x.
\end{equation}
Note that Latin indices denote the boundary coordinates.

We perform a transformation
\begin{eqnarray}
x&=&-1+\frac{2R^2\cos^2\!\theta}{r^2-(m-b^2)\cos^2\!\theta},\\
y&=&-1-\frac{2R^2\sin^2\!\theta}{r^2-(m-b^2)\cos^2\!\theta},\\
\psi&=&\!\frac{\Delta\psi}{2\pi}\,\tilde{\psi}\hspace{-0.09mm}\equiv k_1\tilde{\psi},\\
\phi&=&\frac{\Delta\phi}{2\pi}\,\tilde{\phi}\equiv k_2\tilde{\phi},
\end{eqnarray}
with
\begin{equation}
m=\frac{2(R^2+a^2)}{1-\lambda+a^2/R^2}, \qquad b=\frac{2a}{\sqrt{1-\lambda+a^2/R^2}}.
\end{equation}
Then $0\leqslant\tilde{\psi},\, \tilde{\phi}<2\pi$. In these new coordinates, metric (1) has
the asymptotic form
\begin{eqnarray}
g_{tt}&=&-1\!+\!\frac{2R^2\lambda}{\,1-\lambda+a^2/R^2\,}\frac{1}{r^2}+O(1/r^4),\\
g_{t\tilde{\phi}}&=&-\frac{4R^2\lambda ak_2\cos^2\!\theta}{\,1-\lambda+a^2/R^2\,}\frac{1}{r^2}+O(1/r^4),
\end{eqnarray}
\begin{eqnarray}
g_{rr}&=&1\!+\!\frac{1}{\,1-\lambda+a^2/R^2\,}\Big[4R^2\big(1-\lambda\big)\!\cos^2\!\theta\nonumber\\
&+&\!\big(3R^2\lambda\!-\!R^2\!-\!a^2\big)\Big]\frac{1}{r^2}+O(1/r^4),\\
g_{\theta\theta}&=&r^2\!-\!\frac{2}{\,1-\lambda+a^2/R^2\,}\Big[\big(R^2\!-\!R^2\lambda\!+\!3a^2\big)\!\cos^2\!\theta\!-\!2a^2\Big]+O(1/r^2),\\
g_{r\theta}&=&\frac{3}{2}R^2\sin\!2\theta\frac{1}{r}\!+\!\frac{R^2}{\,2(1-\lambda+a^2/R^2)\,}\Big[2\big(3R^2\!-\!2R^2\lambda\!-\!4a^2\big)\!\cos\!2\theta\nonumber\\
&+&\!3\big(R^2\!+\!R^2\lambda\!-\!a^2\big)\Big]\!\sin\!2\theta\frac{1}{r^3}+O(1/r^5),\\
g_{\tilde{\psi}\tilde{\psi}}&=&\big(1-\lambda+a^2/R^2\big)k^2_1\,r^2\!\sin^2\!\theta+k^2_1\Big[3\big(R^2\lambda\!-\!R^2\!-\!a^2\big)\!\cos^2\!\theta\nonumber\\
&+&\!\big(R^2\!-\!R^2\lambda\!+\!a^2\big)\Big]\!\sin^2\!\theta+O(1/r^2),\\
g_{\tilde{\phi}\tilde{\phi}}&=&\big(1-\lambda+a^2/R^2\big)k^2_2\,r^2\!\cos^2\!\theta+k^2_2\Big[3\big(R^2\lambda\!-\!R^2\!-\!a^2\big)\!\cos^2\!\theta\nonumber\\
&+&\!4a^2\Big]\!\cos^2\!\theta+O(1/r^2).
\end{eqnarray}
Non-vanishing stress-energy tensor components are given by
\begin{eqnarray}
\tau_{tt}&=&\frac{1}{8\pi}\bigg(\!\!\!-\!\frac{R^2}{3\big(1-\lambda+a^2/R^2\big)}\!\Big[20\big(1-\lambda\big)\!\cos\!2\theta\!+\!9\lambda\Big]\frac{1}{r^3}\!+\!O(1/r^5)\bigg),\\
\tau_{t\tilde{\phi}}=\tau_{\tilde{\phi}t}&=&\frac{1}{8\pi}\bigg(\frac{8R^2\lambda ak_2\cos^2\!\theta}{\,1-\lambda+a^2/R^2\,}\frac{1}{r^3}\!+\!O(1/r^5)\bigg),\\
\tau_{\theta\theta}&=&\frac{1}{8\pi}\bigg(\frac{8R^2\big(1-\lambda\big)\cos\!2\theta}{\,3\big(1-\lambda+a^2/R^2\big)\,}\frac{1}{r}\!+\!O(1/r^3)\bigg),\\
\tau_{\tilde{\psi}\tilde{\psi}}&=&\frac{1}{8\pi}\bigg(\frac{8}{3}R^2\big(1-\lambda\big)k^2_1\big(\!\!-\!1\!+\!2\cos\!2\theta\big)\!\sin^2\!\theta\frac{1}{r}\!+\!O(1/r^3)\bigg),\\
\tau_{\tilde{\phi}\tilde{\phi}}&=&\frac{1}{8\pi}\bigg(\frac{8}{3}R^2\big(1-\lambda\big)k^2_2\big(1\!+\!2\cos\!2\theta\big)\!\cos^2\!\theta\frac{1}{r}\!+\!O(1/r^3)\bigg).
\end{eqnarray}

Now we introduce the conservation law
\begin{equation}
D^{i}\tau_{ij}=0,
\end{equation}
where $D^i$ is the covariant derivative adapted to
\vspace{0.7mm}
\begin{equation}
ds^2=-dt^2\!+r^2\big(d\theta^2\!+\sin^2\!\theta d\tilde{\psi}^2\!+\cos^2\!\theta d\tilde{\phi}^2\big).
\end{equation}
Substituting (21)--(25) into (26) yields
\begin{eqnarray}
8\pi D^i\tau_{i\theta}\!=\!\frac{8}{3}R^2(1-\lambda)\!\!\!&\Bigg[&\!\!\!\bigg(\frac{8}{1-\lambda+a^2/R^2}\!-\!4k^2_1\!-\!4k^2_2\bigg)\!\cos^4\!\theta\nonumber\\
&+&\!\bigg(\!\!-\!\frac{8}{1-\lambda+a^2/R^2}\!+\!3k^2_1\!+\!5k^2_2\bigg)\!\cos^2\!\theta\nonumber\\
&+&\!\bigg(\frac{1}{1-\lambda+a^2/R^2}\!-\!k^2_2\bigg)\Bigg]\frac{1}{\sin\!\theta\cos\!\theta}\frac{1}{r^3}\!+\!O\big(1/r^5\big)\!=\!0.
\end{eqnarray}
The coefficient of $r^{-3}$ is identically equal to zero if and only if\hspace{0.6mm}\footnote{When $a=0$ metric (1) describes a static black ring.
To keep the mass of this static ring finite, we must require $\lambda\neq1$. See (2.10) in Ref. 14.}
\begin{eqnarray}
&&\frac{8}{1-\lambda+a^2/R^2}\!-\!4k^2_1\!-\!4k^2_2\!=\!-\frac{8}{1-\lambda+a^2/R^2}\!+\!3k^2_1\!+\!5k^2_2\!=\!\frac{1}{1-\lambda+a^2/R^2}\!-\!k^2_2\!=\!0,\nonumber\\
&&
\end{eqnarray}
namely,
\begin{equation}
k_1=k_2=\frac{1}{\sqrt{1-\lambda+a^2/R^2}}.
\end{equation}
Recall that $k_1\!\equiv\!\frac{\Delta\psi}{\,\,2\pi\,\,}$ and $k_2\!\equiv\!\frac{\Delta\phi}{\,\,2\pi\,\,}$. Then we have
\begin{equation}
\Delta\psi=\Delta\phi=\frac{2\pi}{\sqrt{1-\lambda+a^2/R^2}},
\end{equation}
which is exactly the condition for removing the conical singularity at spatial infinity.
We also use (26) to test the balance of a $\phi$-spinning ring. Setting $k_1\!=\!\frac{1}{\sqrt{1-\lambda+a^2\!/R^2}}$ and
$k_2\!=\!\frac{1}{\sqrt{1+\lambda+a^2\!/R^2}}$,
\begin{equation}
8\pi D^i\tau_{i\theta}
\!=\!\frac{8}{3}R^2(1-\lambda)\bigg(\!\frac{1}{1+\lambda+a^2/R^2}-\frac{1}{1-\lambda+a^2/R^2}\!\bigg)\big(1+2\cos\!2\theta\big)\!\tan\!\theta\frac{1}{r^3}+O(1/r^5)
\end{equation}
is not identically vanishing. This indicates the impossibility of removing both conical singularities in the bulk
and in the boundary, so that a $\phi$-spinning ring can never be in equilibrium.

Based on the above computations, one can go further to study various thermodynamic properties as done in Refs. 9--11.

\section{Discussions}

Finally, it would be interesting to discuss how to bring a $\phi$-spinning ring into equilibrium.
We charge up\cite{16} the vacuum solution (1) in Einstein-Maxwell-dilaton system and obtain
\begin{eqnarray}
&&\hspace{0.3mm}ds^2=-\frac{H(y,x)}{H(x,y)}V_\beta(x,y)^{-2/3}\bigg[dt-\frac{(\lambda a\hspace{-0.2mm}\cosh\!\beta)\hspace{0.2mm}y\hspace{0.2mm}(1-x^2)}{H(y,x)}d\phi\bigg]^{\!2}\nonumber\\
&&\qquad\quad\!+\frac{R^2}{(x-y)^2}H(x,y)V_\beta(x,y)^{1/3}\Bigg[-\frac{dy^2}{(1-y^2)F(y)}-\frac{(1-y^2)F(x)}{H(x,y)}d\psi^2\nonumber\\
&&\qquad\qquad\qquad\quad\ \ \,\qquad\qquad\qquad\qquad\,\hspace{0.36mm}+\,\frac{dx^2}{(1-x^2)F(x)}+\frac{(1-x^2)F(y)}{H(y,x)}d\phi^2\Bigg],\nonumber\\
&&\\
&&\hspace{0.1mm}A=\frac{1}{2}\frac{\sinh\!\beta}{H(x,y)V_\beta(x,y)}\bigg[\Big(H(x,y)\!-\!H(y,x)\Big)\!\cosh\!\beta\hspace{0.5mm}dt+\lambda a y\big(1-x^2\big)d\phi\bigg],
\end{eqnarray}
\begin{equation}
\hspace{-8.39cm}e^{-\Phi}=V_\beta(x,y)^{\frac{1}{\sqrt{6}}},
\end{equation}
where $A$ is the gauge potential, $e^{-\Phi}$ is called the dilaton, $F(\xi)$ as well as $H(\xi_1,\xi_2)$ are
the same as those in (2), and
\begin{equation}
V_\beta(x,y)=\cosh^2\!\beta-\sinh^2\!\beta\frac{H(y,x)}{H(x,y)}.
\end{equation}
One can see that a new parameter $\beta$ associated with the charge is added. (33)--(35) satisfy
the field equations
\begin{equation}
\mathcal{R}^{^{(5)}}_{\mu\nu}=2\partial_\mu\Phi\hspace{0.3mm}\partial_\nu\Phi+2e^{-2\alpha\Phi}\big(F_{\mu\rho}F^{\,\,\,\rho}_\nu\!-\frac{1}{6}g_{\mu\nu}F_{\rho\sigma}F^{\rho\sigma}\big),
\end{equation}
\begin{equation}
\nabla_\mu\big(e^{-2\alpha\Phi}F^{\mu\nu}\big)\!=0,
\end{equation}
\begin{equation}
\nabla_\mu\!\nabla^{\mu}\Phi+\frac{\alpha}{2}e^{-2\alpha\Phi}F_{\rho\sigma}F^{\rho\sigma}\!=0,
\end{equation}
where $\alpha\!=\!\sqrt{8/3}$ is the Kaluza-Klein coupling constant. To balance the tension and
gravitational self-attraction, we should immerse the resulting charged ring in an external
electric field by performing the Harrison transformation (see Ref. 17). However, since the
gauge potential (34) has more than one component, the familiar ``standard" Harrison transformation
does not work. This would probably require techniques along the lines of Refs. 18--20. On the
other hand, we also hope to construct the $\phi$-spinning ring carrying dipole charges. Typically,
the gauge potential of a dipole ring has only one component and thus allows application of
the Harrison transformation. We will focus on these aspects in the future.

\section*{Acknowledgments}

This work was supported by NSFC grant No. 10775175.


\end{document}